\documentclass[preprint,12pt,3p]{elsarticle}

\usepackage{graphics}
\usepackage{graphicx}
\usepackage{epsfig}

\usepackage{amssymb}
\usepackage{xcolor}
\usepackage{times}
\usepackage{epsfig}
\usepackage{graphicx}
\usepackage{subfig}
\usepackage{float}
\usepackage{amsmath}
\usepackage{amssymb}
\usepackage{algorithm}
\usepackage{algorithmic}
\usepackage{multirow}
\usepackage{setspace}
\makeatletter \@fpsep = 2pt \makeatother
\begin{document}

\begin{spacing}{1.0}

\begin{frontmatter}

\title{Lightweight Image Super-Resolution with Enhanced CNN}

\author[label1,label2]{Chunwei Tian}
\author[label1,label2]{Ruibin Zhuge}
\author[label1,label2]{Zhihao Wu}
\author[label1,label2,label3]{Yong Xu\corref{cor1}}
\ead{yongxu@ymail.com}
\cortext[cor1]{Corresponding author}
\author[label4,label3]{Wangmeng Zuo}
\author[label5]{Chen Chen}
\author[label6]{Chia-Wen Lin}
\address[label1]{Bio-Computing Research Center,  Harbin Institute of Technology, Shenzhen, Shenzhen, 518055, Guangdong, China}
\address[label2]{Shenzhen Key Laboratory of Visual  Object  Detection  and  Recognition,  Shenzhen,  518055,  Guangdong, China.}
\address[label3]{Peng Cheng Laboratory, Shenzhen, 518055, Guangdong, China}
\address[label4]{School of Computer Science and Technology, Harbin Institute of Technology, Harbin, 150001,  Heilongjiang, China}
\address[label5]{Department of Electrical and Computer Engineering, University of North Carolina at Charlotte, NC, 28223, USA}
\address[label6]{Department of Electrical Engineering and the Institute of Communications Engineering, National Tsing Hua University,
Hsinchu, Taiwan}
% \ead{youxg@hust.edu.cn}
% \ead{ymc@comp.hkbu.edu.hk}

\begin{abstract}
Deep convolutional neural networks (CNNs) with strong expressive ability have achieved impressive performances on single image super-resolution (SISR). However, their excessive amounts of convolutions and parameters usually consume high computational cost and more memory storage for training a SR model, which limits their applications to SR with  resource-constrained devices in real world. To resolve these problems, we propose a lightweight enhanced SR CNN (LESRCNN) with three successive sub-blocks, an information extraction and enhancement block (IEEB), a reconstruction block (RB) and an information refinement block (IRB). Specifically, the IEEB extracts hierarchical low-resolution (LR) features and aggregates the obtained features step-by-step to increase the memory ability of the shallow layers on deep layers for SISR. To remove redundant information obtained, a heterogeneous architecture is adopted in the IEEB. After that, the RB converts low-frequency features into high-frequency features by fusing global and local features, which is complementary with the IEEB in tackling the long-term dependency problem. Finally, the IRB uses coarse high-frequency features from the RB to learn more accurate SR features and construct a SR image. The proposed LESRCNN can obtain a high-quality image by a model for different scales.  Extensive experiments demonstrate that the proposed LESRCNN outperforms state-of-the-arts on SISR in terms of qualitative and quantitative evaluation. The code of LESRCNN is accessible on https://github.com/hellloxiaotian/LESRCNN.
\end{abstract}


\end{frontmatter}

% \linenumbers

\section{Introduction}
\label{sec-1}
Single image super-resolution (SISR) aims at recovering a high-resolution (HR) image from a low-resolution (LR) observation. Since multiple HR images can be downsampled to the same LR image, SISR is an ill-posed problem \cite{zhang2019deep}. For solving this issue, prior knowledge methods were developed by constraining the solution space \cite{xu2017multi,zhao2017image}. For instance, the sparse-coding method in \cite{zha2020benchmark} combines LR patches and dictionary learning to obtain  super-resolution (SR) patches, and then applies weighted averaging to produce the high-resolution (HR) image \cite{yang2010image}. The non-local means with steering kernel regression method in \cite{zhang2012single} jointly utilizes non-local and local priors to extract complementary information for enhancing SR performance. To accelerate the training of a SR model, it was proposed in \cite{zuo2011generalized} to generalize the original accelerated proximal gradient to take the place of the Lipschitz constant to improve the convergence process. In addition, there have been some good tools, such as random forest \cite{schulter2015fast} and regression \cite{timofte2014a+} proposed for SISR. However, these methods rely on external example information to improve the performance of SISR, which leads to two drawbacks significantly limiting their applications. First, most of these methods resort to complex optimization methods to enhance the qualities of recovered HR images and other low-level tasks at the expense of efficiency \cite{ren2019simultaneous,li2020robust}. Second, they usually require manually tuning parameters to boost the SR performance.

To address the above problems, various convolutional neural networks (CNNs) with flexible end-to-end network architectures and effective training strategies were proposed \cite{tian2020image,yuan2020learning,tian2020designing},  having brought prosperous development in image restoration tasks, especially image super-resolution \cite{yang2019deep}. Dong et al. \cite{dong2015image} proposed a pioneering three-layer SRCNN to obtain the SR image in a pixel mapping manner. Although the shallow SRCNN was simpler and more effective than traditional SR techniques, it was hard to make a tradeoff among depth, effectiveness and performance. Since then, the designs of deeper networks which pursue superior SR performance have become popular. For example, the cascade of sparse coding based on networks (CSCN) in \cite{wang2015deep} utilize a sparse coding technique to guide a deep network for accelerating the training speed and compressing the SR model. A very deep SR network (VDSR) was proposed in \cite{kim2016accurate} to enlarge dramatically the depth of the network by stacking multiple layers to enhance SR performance. To prevent vanishing or exploding gradients, skip connections and recursive operations in deep networks were proposed \cite{kim2016deeply}. The deep recursive residual network (DRRN) in \cite{tai2017image} uses recursive learning to control the number of parameters. Besides, global and local residual learning (RL) techniques are incorporated into DRRN to facilitate the training for SISR. A very deep persistent memory network (MemNet) was proposed in  \cite{tai2017memnet} that applies multiple recursive units and gate units to extract and fuse multi-level features to enhance the visual qualities of reconstructed HR images. The 60-layer residual encoder-decoder network (RED30) in \cite{mao2016image} employs a symmetric network architecture via skip connections to effectively extract details of a HR image. Although most of the above-mentioned methods can boost the visual qualities of SR results, the inputs of these networks are first upsampled to the same resolution as the output sizes for training a SR model, thereby increasing the computational cost and memory consumption significantly \cite{dong2016accelerating}.

To better trade SR performance for resource consumption, the fast SR CNN (FSRCNN) in \cite{dong2016accelerating} utilizes sub-pixel convolution as the final layer to upscale the resolution of obtained feature map, thereby speeding up the SR process while degrading visual quality. To address this issue, novel network architecture based on image characteristics and multi-level feature integration have been attracting increasing attention. For example, the enhanced deep SR network (EDSR) \cite{lim2017enhanced} exploits improved ResNet architecture in \cite{he2016deep} and multi-scale techniques to gain performance improvement  in image SR. The residual dense network (RDN) in \cite{zhang2018residual} based on EDSR utilizes global and local features to enhance the diversity of the network architecture for improving the SR performance. The multi-level wavelet CNN (MWCNN) in \cite{liu2018multi} incorporates signal processing technique into the U-Net to promote the performance and computational efficiency for restoration tasks.

Most of these algorithms above resort to increasing the depth of a SR network to enhance the expressive power of the network for performance improvement. However, this usually results in more parameters and excessive memory consumption, which is usually not affordable for resource-constrained mobile devices in practical applications.

In this paper, we propose a lightweight enhanced super-resolution CNN (LESRCNN) by cascading three sub-blocks, an information extraction and enhancement block (IEEB), a reconstruction block (RB), and an information refinement block (IRB). The IEEB uses hierarchical LR features and residual learning techniques to enhance the memory ability of  shallow layers for improving the SR performance. By incorporating the heterogeneous architecture proposed in \cite{singh2019hetconv} into the IEEB, the amounts of parameters and memory consumption for the IEEB are significantly reduced, so as the training time. Then, the RB fuses the extracted global and local features to transform low-frequency features (i.e., the LR features) into high-frequency features (i.e., the HR features) via residual learning and sub-pixel convolution methods. This also leads to an auxiliary effect of preventing the long-term dependency problem with the IEEB. Finally, the IRB uses the coarse high-frequency features from the RB to learn more accurate SR features and construct a SR image.

The contributions of the proposed LESRCNN are summarized as follows.

(1) LESRCNN significantly reduces the number of parameters for achieving excellent performance on SISR by cascading an information extraction and enhancement block, a reconstruction block and an information refinement block. As a result,  the low computational cost and memory consumption make LESRCNN particularly suitable for resource-constrained mobile devices for real-world applications.

(2) The information extraction and enhancement block extracts hierarchical LR features and fuses them to enhance the memory ability of shallow layers for improving the SISR performance. Besides, we also propose a heterogeneous architecture in the  information extraction and enhancement block for compressing the network, which significantly reduces the computational cost and memory consumption. Moreover, since LR patches are used to train  the SR model, the training process can be significantly accelerated.

(3) The reconstruction block combines global and local features by residual learning and sub-pixel convolution techniques to convert low-frequency features into high-frequency features, which is complementary with the   information extraction and enhancement block in preventing the long-term dependency problem.

(4) The information refinement block applies coarse high-frequency features extracted by the reconstruction block to learn more accurate high-frequency features to effectively enhance the fidelity of the  predicted SR image with respect to its HR ground-truth. The proposed method can deal with different scales via a model for SISR.

The remainder of this paper is organized as follows. Section 2 presents related work. Section 3 illustrates the proposed method. Section 4 provides extensive experimental results and Section 5 concludes the paper.

\section{Related Work}
\subsection{Deep CNNs based cascaded structures for SISR}
With the rapid development of big data and graphic processing unit (GPU), deep CNNs have widely applied in SISR. The SR techniques based on deep CNNs mainly consist of three kinds: using high-frequency features for training a SR model, using low-frequency features for training a SR model and combination of high- and low-frequency features for training a SR model. The first method, such as DRRN \cite{tai2017image}, MemNet \cite{tai2017memnet} and RED \cite{mao2016image} upsamples the LR image the same as the given HR image as the input of deep SR network, which causes higher computational cost and more memory consumption. The second method, i.e., FSRCNN \cite{dong2016accelerating} only used sub-pixel convolution technique as the final layer in the SR network to amplify the extracted low-frequency features, which ignored the detailed information from high-frequency features. Although this method is superior to training speed, their SR performance is unsatisfactory. The third method simultaneously uses high- and low-frequency features to recover the high-quality image, which is very popular in SISR. Specifically, deep CNNs based cascaded structures can better express the third method above. Deep CNNs based cascaded structures can be divided into two categories from measuring SR effects: performance and efficiency.

In improving the SISR performance, cascading multistage networks can improve the resolution step by step \cite{ahn2018image}. A coarse-to-fine CNN \cite{tian2020coarse} uses heterogeneous convolutions in a stack of feature extraction blocks to extract low-frequency features, then, applies feature refinement block to learn more accurate high-frequency features for image-resolution. A cascading dense network (CDN) \cite{wei2019accurate} can extract hierarchical features from each convolutional layer, then, densely connect these obtained features to eliminate vanishing gradient and enhance the SR performance. Enlarging the width of network can urge more robust features in SISR. Thus, cascading two sub-networks to enlarge the width was a good choice to improve the expressive ability of the SISR model \cite{chowdhury2019single}.

In improving the efficiency of training a SR model, compressing deep networks is very effective. A cascading residual network (CARN) \cite{ahn2018fast} used multiple cascading connections to gather recursive blocks for guaranteeing performance of the SR model. Also, convolutions of size $1 \times 1$ were fused into the CARN to reduce the number of parameters and cut down the training time. Channels were divided into groups to simultaneously learn new features, which can improve the efficiency of training on SISR. The group convolutions and weight-typing were fed into a residual network \cite{ahn2019photo} to obtain extreme efficiency for dealing with a LR image. Inspired by the facts above, we design a deep network based on cascaded structure to extract accurate LR and HR features for improving the training stability in terms of the SISR task.
\subsection{Deep CNNs based blocks for SISR}
Plug-and-play architectures enlarge the flexibilities of deep CNNs on different computer vision tasks, such as video \cite{yuan2020visual,li2019fast}, text-to-image synthesis \cite{liang2019cpgan}, image denoising \cite{tian2019enhanced}, image deraining \cite{ren2020brn}, low-light image enhancement \cite{ren2019low}, image dehazing \cite{ren2020single} and image super-resolution \cite{zhang2018learning}. Specifically, deep CNNs based blocks can better cooperate with each component to facilitate more useful information, which is popular in real applications. This method can be divided into two groups: pursuing better performance and taking lower computational cost.

For the first aspect, features fusion methods can achieve superior performance against other methods. Gathering the same types of different feature levels of each cascaded sub-network can capture more contexts to achieve the aim of recovered high-quality image \cite{hu2018single}. To address long-term dependency question, different types of different feature levels are merged to improve the discriminative ability for SISR. For example, Hu et al. \cite{hu2019channel} combined channel-wise and spatial features into the cascaded networks to promote the representation capacity in SISR. To deal with difficult training of the deep network, multi-scale technique jointed the dilated convolutional neural network to make obtained features better interact for obtaining more accurate SR image \cite{shi2017single}. Additionally, according to the image nature, enhancing the effects of important features is very effective in low-level vision tasks. Zhang et al. \cite{zhang2018image} presented a channel attention mechanism into the residual network to adjust the influences of useful channels, which obtained better accuracy and visual improvements for SISR.

For the second aspect, compressing the network has become mainstream technology. \cite{wu2020revisiting} uses the knowledge distillation to transfer a general model to more personalized models, which can improve the training efficiency. In addition, the idea of distillation is first proposed by \cite{hinton2015distilling}. \cite{wu2020learning} used the combination of correlated embedding loss and knowledge distillation to resolve image recognition task. Using small convolutions is also popular to compress model. A novel information distillation network (IDN) \cite{hui2018fast} utilized group convolutions and small convolutions of $1 \times 1$ to remove the non-essential parts of deep network, which was useful to reduce the computational cost and complexity for training a SR model. Reducing the dimension of the data can also improve the speed in handling image restoration tasks. For example, Lai et al. \cite{lai2018fast} utilized Laplacian Pyramid network with progressive upsampling to reduce the number of parameters and obtain better performance in image super-resolution. Using signal processing idea or machine learning methods to guide the design of deep network can facilitate more features. Inspired by the fractal structure, Zhang et al. \cite{zhang2020adaptive} applied adaptive importance learning to guide
the CNN for SISR. Additionally, according to importance of candidate features, designing an efficient CNN was very effective. Hui et al. \cite{hui2019lightweight} exploited group convolutions of size $3 \times 3$ and convolutions of $1 \times 1$ to enhance and distill the obtained features, respectively, then, they applied an attention module to enhance the importance of key features, which obtained competitive result and fast execution ability.

According to previous researches, we can see that these methods applied different mechanisms to deal with SISR task. However, designs of their network architectures broke the rules of improving the performance or reducing computational resource. Motivated by that, we utilize deep CNNs based blocks to make a tradeoff between performance and computational cost for SISR, which is suitable to real applications.

\section{Proposed Method}
Our proposed LESRCNN is described in Figs. 1 and 2. LESRCNN is implemented by cascading an information extraction and enhancement block (IEEB), a reconstruction block (RB) and an information refinement block (IRB). The IEEB extracts hierarchical low-frequency features and aggregates these features step-by-step by residual learning to increase the memory ability of shallow layers on deep layers. This can enhance the SR performance without significantly increasing computational cost. Also, to remove redundant information obtained, we propose a heterogeneous architecture in the IEEB. After that, the RB transforms the extracted low-frequency features into high-frequency features by fusing the global and local features. This complements of the information extraction and enhancement block can address the long-term dependency problem. Finally, the IRB  refines  the coarse high-frequency features to derive more accurate SR features and obtain a SR image. These blocks are explained in details in later subsections.
\begin{figure}[!htb]
\centering
\subfloat{\includegraphics[width=6.5in]{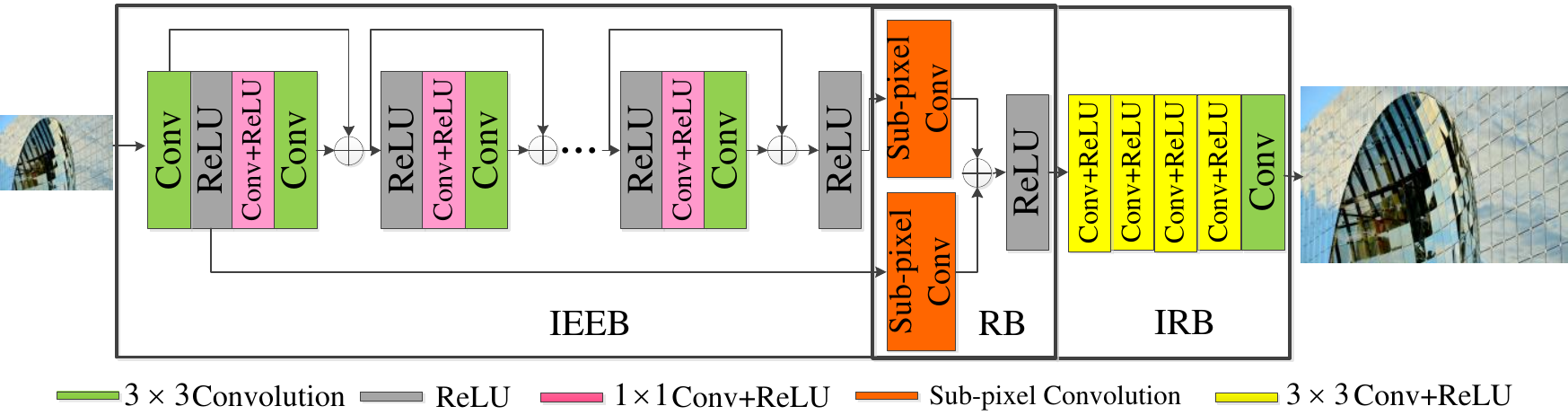}
\label{fig_second_case}}
\caption{Network architecture of the proposed LESRCNN.}
\label{fig:5}
\end{figure}
\begin{figure}[!htb]
\centering
\subfloat{\includegraphics[width=5.5in]{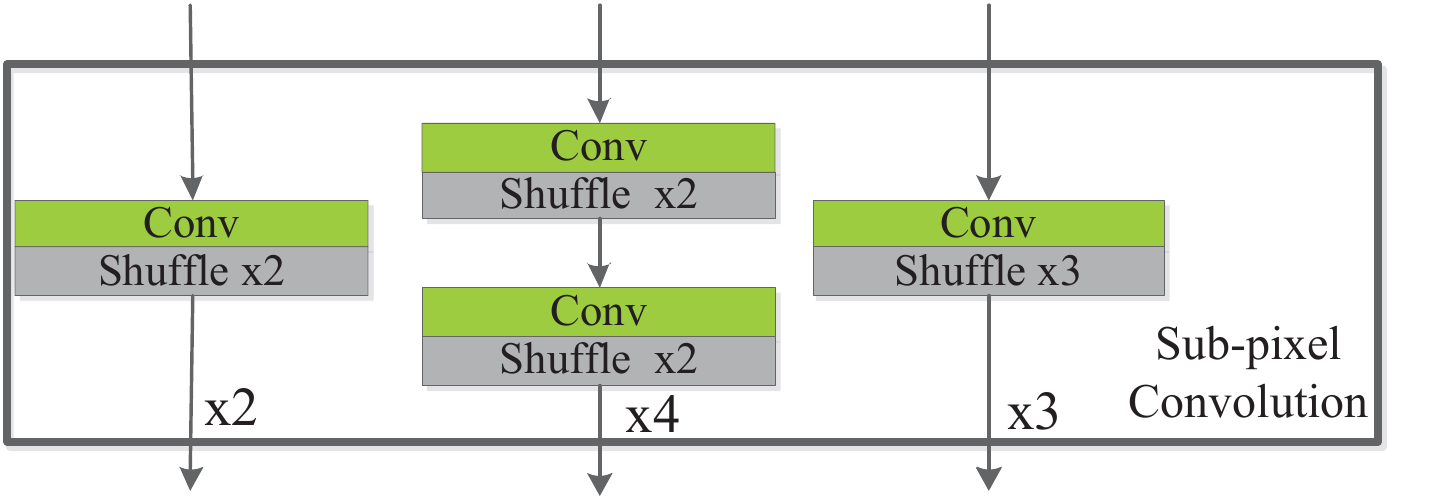}
\label{fig_second_case}}
\caption{Components of the sub-pixel convolution.}
\label{fig:5}
\end{figure}
\subsection{Network architecture}
The proposed 23-layer LESRCNN consists of three blocks, an IEEB, a RB and an IRB. The 17-layer IEEB extracts and enhances the low-frequency features, and then refines the extracted low-frequency features to reduce computation. Then, the 1-layer RB converts these low-frequency features to high-frequency features.
Finally, the 5-layer IRB refines the coarse high-frequency features extracted by the RB to derive more accurate SR features, which is useful to enhance the fidelity between the predicted SR and its HR ground-truth. To better explain these modules. We define the following terms. Let ${I_{LR}}$ and ${I_{SR}}$ represent the input LR image and the recovered SR image, respectively, ${f_{IEEB}}$, ${f_{RB}}$ and ${f_{IRB}}$ denote the functions of the IEEB, the RB and the IRB, respectively. The SR process with LESRCNN can be formulated as follows:
\begin{equation}\label{2}
\begin{split}
{O_{SR}} = &{f_{IRB}}({f_{RB}}({f_{IEEB}}({I_{LR}}))),\\
{\rm{       =  }}&{f_{LESRCNN}}({I_{LR}})\\
\end{split}
\end{equation}
where ${f_{LESRCNN}}$ denotes the function of LESRCNN. The SR performance of LESRCNN relies on an appropriately defined loss function, as will be elaborated in Section 3.2.
\subsection{Loss function}
We use mean squared error (MSE) \cite{douillard1995iterative} as the metric to measure the discrepancy between a reconstructed SR image with its HR ground-truth. We use a set of training image pairs $\{ I_{LR}^i,I_{HR}^i\} _{i = 1}^T$ to calculate the MSE, where $T$ is the total number of training images. $I_{LR}^i$ and $I_{HR}^i$ are the $i$-th LR and HR images, respectively. To train LESRCNN, we aim at minimizing the following loss function:
\begin{equation}\label{2}
\begin{split}
l(p ) = \frac{1}{{2T}}{\sum\limits_{i = 1}^T {\left\| {{f_{LESRCNN}}(I_{LR}^i) - I_{HR}^i} \right\|} ^2},
\end{split}
\end{equation}
where $p$ is the parameter set of the SR model.
\subsection{Information extraction and enhancement block (IEEB)}
Typically, the deeper the depth of a network, the poorer the memory ability of the shallow layers. To handle this issue, we propose a 17-layer information extraction and enhancement block (IEEB) to achieve both good performance and high efficiency. IEEB extracts hierarchical LR features, and then uses residual learning to fuse the hierarchical features to preserve the effects of shallow-layer features on deep layers. Meanwhile, IEEB utilizes a heterogeneous architecture  to distill the obtained features so as to reduce the number of parameters, computational cost, and memory consumption. The 17-layer IEEB involves two types of convolution: $3 \times 3$ Conv+ReLU and $1 \times 1$ Conv+ReLU, where Conv+ReLU represents a convolution followed by a ReLU activation function \cite{krizhevsky2012imagenet}. Specifically, the odd layers of 1, 3, 5, , ..., 17 are $3 \times 3$ Conv+ReLU. Note, the size of the first layer is $3 \times 3\times 3\times 64$, where $3$, $3 \times 3$ and $64$ denote the channel number of the input, filter and channel number of output, respectively. The sizes of other odd layers are $64 \times 3\times 3\times 64$, where $64$, $3 \times 3$ and $64$ are the channel of the input, filter and channel of output, respectively. To preserve the information of shallow layers, we fuse the hierarchical information via the residual learning method. That is, the current odd layer has effect on itself and the following odd layers. For example, the first-layer features work for layers 1, 3, 5, ..., 17. Moreover, to reduce the computational cost and memory consumption, we set the convolution used in the even layers (i.e., layers 2, 4, 6, ..., 16) as $1 \times 1$ Conv+ReLU.  Let ${C_3}$ and ${C_1}$ denote the convolutional functions with sizes of $3 \times 3$ and $1 \times 1$ , respectively. $R$  the function of the ReLU. $O_c^i$  the output of the convolution of the $i$-th layer, and ${O_i}$ the output of the $i$-th layer, where $i = 1,2,...,,17$. Specifically, $O_c^1 = {C_3}({I_{SR}})$ and ${O_1} = R(O_c^1)$. The outputs of the subsequent convolutional layers are as follows.
\begin{equation}\label{2}
O_c^i=\left\{
\begin{array}{ll}
{C_3}({O_{i - 1}})     &      \text{$i$ is odd}\\
{C_1}({O_{i - 1}})   &      \text{$i$ is even}\\
\end{array} \right.,
\end{equation}
where  $i \in (2,17)$ is the the layer index. According to the explanations above and Fig. 1, the output of each layer is formulated as
\begin{equation}\label{2}
O_j=\left\{
\begin{array}{ll}
R(O_c^j + \sum\limits_{j = 1}^{j - 2} {O_c^j} )     &      \text{$j$ is odd}\\
R(O_c^j){\rm{              }}   &      \text{$j$ is even}\\
\end{array} \right.,
\end{equation}
where odd $j = 3,5,7,...,17$. If $j$ is even, its value falls in $2,4,6,...,16$. Further, the output of IEEB is presented as ${O_{17}} = R(O_c^{17} + \sum\limits_{i = 1}^{15} {O_c^j} )$, where `+' denotes the residual learning technique and $\oplus$ is used to represent `+' in Fig. 1. Additionally, the output of the IEEB, ${O_{17}}$ acts the RB.
\subsection{Reconstruction block (RB)}
Many existing methods use bicubic interpolation to upscale the input LR image to the target size as the input of a SR model, which, however, consumes significantly more computation and memory  \cite{dong2016accelerating}. To address this issue, we incorporate  the 1-layer sub-pixel convolution proposed in \cite{shi2016real} into the reconstruction block (RB)  to convert low-frequency features to high-frequency ones, where the convolution filter size is $3 \times 3$ and its channels of input and output are both $64$. The sub-pixel convolution used in LESRCNN is divided into two kinds: a model for three scales and a model for one scale. When a SR model (i.e., LESRCNN-S) is trained for three scales, the sub-pixel convolution is composed of three components: Conv+Shuffle $\times 2$, two Conv+Shuffle $\times 2$, Conv+Shuffle $\times 3$, where Conv+Shuffle $\times 2$ denotes convolution of $3 \times 3$ connects Shuffle $\times 2$ and  Conv+Shuffle $\times 3$ is convolution of $3 \times 3$ connects Shuffle $\times 3$. The Conv+Shuffle $\times 2$ and two Conv+Shuffle $\times 2$ are used for $\times 2$ and $\times 4$, respectively. The Conv+Shuffle $\times 3$ is applied for $\times 3$. When a certain SR model (i.e., LESRCNN) is trained for single scale, the sub-pixel convolution technique only has a component from Conv+Shuffle $\times 2$, two Conv+Shuffle $\times 2$ and Conv+Shuffle $\times 3$, as shown in Fig. 2.

In addition, to further address long-term dependency problem, we integrate global and local features to enhance the memory ability of shallow-layer features in deep layers. The function of RB is summarized in the following two steps. The first step uses the sub-pixel convolution  to upsample the outputs of the 1st and 16th layers as global and local features, respectively. The second step utilizes residual learning to fuse the global and local features for enhancing the SR performance. After that, we apply the ReLU to convert the result of the second step into non-linearity. The process of RB can be formulated as
\begin{equation}\label{2}
{O_{RB}} = R(S({O_1}) + S({O_{17}})),
\end{equation}
where $S(\cdot)$ denotes the sub-pixel convolution, ${O_1}$ and ${O_{17}}$ represent the global and local features, respectively, and ${O_{RB}}$ represents the output of the RB.
\subsection{Information refinement block (IRB)}
As explained previously, combining the high- and low-frequency features to recover high-quality image is effective for SISR. However, the proposed IEEB only uses the LR image to extract low-frequency features. RB is then employed to convert the obtained low-frequency features to coarse high-frequency features, which may lack detailed information of high-frequency features. To address this problem, we propose an information refinement block (IRB) to learn more accuracy SR features and reconstruct the final SR image accordingly.   The 5-layer IRB consists of 4-layer Conv+ReLU and 1-layer Conv. The Conv+ReLU layer contains  a convolutional followed by a ReLU, where filter size of convolutional layer is $3 \times 3$ and the channel numbers of input and output are both $64$. The filter size of the final convolutional layer is $3 \times 3$, and the channel numbers of input and output are $64$ and $3$, respectively. The function of IRB is formulated as.
\begin{equation}\label{2}
{O_{SR}} = {C_3}(R({C_3}(R({C_3}(R({C_3}(R({C_3}({O_{RB}})))))))))
\end{equation}
\section{Experiments}
\subsection{Training dataset}
By following \cite{ahn2018fast}, the public DIV2K dataset \cite{agustsson2017ntire} is used as training dataset for a SR model in this paper. The DIV2K dataset comprises 800 training images, 100 validation images and 100 test images under different scales of $\times 2$, $\times 3$ and $\times 4$, where they are saved in the format of `.png'. Specifically, enlarging the dataset is useful to improve the performance in image applications \cite{krizhevsky2012imagenet}. Motivated by that, we merge the training and validation datasets to form novel training dataset. Additionally, to reduce the training cost, each LR image is cropped as patches with size $64 \times 64$, which can improve the efficiency \cite{xu2015patch} of training. Further, random horizontal flips and $90^\circ$ rotation operations are applied to augment obtained training patches.
\subsection{Test datasets}
For the test phase, four benchmark datasets \cite{yang2014single}, such as Set5 \cite{bevilacqua2012low}, Set14 \cite{yang2010image}, BSD100 \cite{martin2001database} and Urban100 \cite{huang2015single} with different scales of $\times 2$, $\times 3$ and $\times 4$ are chosen, which are saved in the format of `.png'. The Set5 and Set14 have five and fourteen color images with different background, respectively. The 100 color images with different scenes are captured in the BSD100 (also treated as B100) and Urban100 (also named U100), respectively.

It is known that the SR methods (i.e., RED \cite{mao2016image}) employed Y channel of YCbCr space to design experiments. Following the rule, the predicted RGB image from the LESRCNN is transformed into the Y channel to test the performance of SISR in this paper.
\subsection{Implementation details}
During the training, we set the initial parameters as follows. Batch size and epsilon are 64 and 1e-8, respectively. Beta\_1 and beta\_2 are 0.9 and 0.999, respectively. The training process has 6e+5 steps. The initial learning rate is set to 1e-4 and halved every 4e+5 steps. Additionally, other initial parameters are the same as \cite{ahn2018fast}. The trained model is updated by Adam optimizer \cite{kingma2014adam}.

The LESRCNN is implemented by Pytorch of 0.41 and Python of 2.7. The related codes run on Ubuntu of 16.04 from a PC, which consists of a CPU of Inter Core i7-7800, a RAM of 16G and two GPUs of Nvidia GeForce GTX 1080Ti in this paper. The two GPUs can be accelerated by Nvidia CUDA of 9.0 and CuDNN of 7.5.
\subsection{Network analysis}
The proposed LESRCNN takes lower computational cost to obtain better performance in SISR. Its implementations by three blocks: an information extraction and enhancement block, a reconstruction block and an information refinement block. The information extraction and enhancement block, IEEB makes full use of hierarchical low-frequency features to enlarge the memory ability of shallow layers on deep layers. Meanwhile, a heterogeneous architecture is fused into the information extraction and enhancement block to distill obtained information, which is beneficial to reduce the computational cost and memory consumption. The reconstruction block converts the obtained low-frequency features from the IEEB into high-frequency features via the sub-pixel convolution technique. Then, it uses the RL method to fuse global and local features for better addressing long-term dependency problem, which can support the IEEB. Finally, the information refinement block is utilized to learn more accurate high-frequency features and construct a SR image. These techniques cooperate well to outperform state-of-the-art SR methods, such as the RED for SISR. Further, the design principles of the mentioned key techniques are shown in details as follows.

(1) Information extraction and enhancement block: In real applications, performance and computational cost are very important to mobile devices \cite{tian2020attention}. Thus, the design of the IEEB breaks the rules of lower computational cost and less memory consumption, and higher performance for SR task. For reducing the training cost, convolution of smaller filter size (e.g. $1 \times 1$) is a good choice to compress the network \cite{hui2018fast}. However, choosing the locations of convolutions of $1 \times 1$ are difficult. Due to the lower computational cost and excellent performance, a heterogeneous architecture \cite{singh2019hetconv} is chosen to address this problem. Specifically, the heterogeneous architecture comprises heterogeneous convolutions. Here heterogeneous convolutions with $P = 2$ \cite{singh2019hetconv} are used in the IEEB, where $P$ represents part. The heterogeneous convolutions with $P = 2$ denote that each standard convolution of $3 \times 3$ and each convolution of $1 \times 1$ are connected. More information of heterogeneous convolutions refers to \cite{singh2019hetconv}. To convert obtained features into non-linearity, the activation function of ReLU is set behind each convolution in this paper. To verify the effects of the heterogeneous convolutions for SISR, we conduct the experiments by heterogeneous convolutional network (HN) and standard convolutional network (SN) in terms of performance, running time and complexity as shown in Tables 1-3. Specifically, the 17-layer HN comprises sixteen heterogeneous convolutions (eight convolutions of $3 \times 3$ and eight convolutions of $1 \times 1$) and a standard convolution of $3 \times 3$, where each convolution connects with a ReLU. It is known that enlarging the diversity of network is useful to promote the performance in image processing tasks \cite{zhang2018residual}. Motivated by the fact, the seventeenth layer of the HN is added. Additionally, the aim of SR task is to obtain the HR image. Thus, we use a sub-pixel technique behind the HN to convert LR features into SR features. And a single convolution of $3 \times 3$ as the final layer of deep network is used to construct a predicted SR image. The SN has the same depth and components as the HN. However, it is noted that the sizes of the convolutions in the SN are $3 \times 3$.

In terms of SR performance, the SN is slightly higher than the HN on Set5 under scale of $\times 4$ for both of peak signal-to-noise ration (PSNR) and structural similarity index (SSIM) \cite{hore2010image} as illustrated in Table 1. However, the SN is higher than the HN in running time of a given LR image as shown in Table 2. Also, the SR has higher cost than that of the HR as presented in Table 3. According to the analysis above, we can see that the HN is more competitive than the SN in performance, running time and training cost for SISR. Thus, the HN fused into the IEEB for real applications, such as mobile device is reasonable.
\begin{table*}[t!]
\caption{Average PSNR and SSIM of different methods under scale of $\times 4$ on Set5.}
\label{tab:1}
\centering
\scalebox{1}[1]{
\begin{tabular}{|c|c|c|}
\hline
\multirow{2}{*} {Scale} & \multirow{2}{*}{Methods} &Set5\\
\cline{3-3} &&PSNR/SSIM \\
\hline
\multirow{5}{*}
{$\times 4$}&SN  &31.64/0.8864  \\
\cline{2-3} &HN  &31.62/0.8852\\
\cline{2-3} &IEEB  &31.73/0.8877\\
\cline{2-3}  &IEEB+RB &31.76/0.8881\\
\cline{2-3}  &LESRCNN &31.88/0.8903\\
\hline
\end{tabular}}
\end{table*}
\begin{table}[t!]
\caption{Running time of two methods for predicting a SR image of sizes $256 \times 256$, $512 \times 512$ and $1024 \times 1024$.}
\label{tab:1}
\centering
\scalebox{1}[1]{
\begin{tabular}{|c|c|c|}
\hline
\multirow{3}{*}{Sizes} &
\multicolumn{2}{c|}{Methods}\\
\cline{2-3}
 &SN &HN\\
\cline{2-3}
 & \multicolumn{2}{c|}{$\times 4$}\\
\hline
$256 \times 256$ &0.00669	&0.00651\\
\hline
$512 \times 512$ &0.00879	&0.00869\\
\hline
$1024 \times 1024$ &0.01672	&0.01651\\
\hline
\end{tabular}}
\label{tab:booktabs}
\end{table}
\begin{table}[t!]
\caption{Complexity of two comparative methods.}
\label{tab:1}
\centering
\scalebox{1}[1]{
\begin{tabular}{|c|c|c|}
\hline
Methods &Parameters &Flops\\
\hline
SN  &630K &3.06G\\
\hline
HN	&368K  &1.38G\\
\hline
\end{tabular}}
\label{tab:booktabs}
\end{table}

It is known that as the growth of depth, memory ability of shallow layers gets weaker, which results in the consequence that the performance of image applications gets poorer \cite{tai2017memnet,tian2019deep}. For resolving this problem, multi-level feature fusion idea is applied in this IEEB. That is, we make full use of hierarchical information without increasing the computational cost to enhance the effects of shallow layers on deep layers in SISR. The detailed information of enhancement operation is given in Section 3.3. Additionally, the enhancement operation also increases the diversity of the network architecture, which is useful to promote the SR performance. These illustrations are proved in Table 1, where the `IEEB' has higher PSNR and SSIM than that of the `HN'. That shows that the enhancement operation is very effective for SISR. And the design of the IEEB is rational and effective in recovering a HR image.

(2) Reconstruction block: It is indisputable fact that employing bicubic interpolation to upsample the original LR image can bring greater training cost \cite{dong2016accelerating}. For tackling the problem, the sub-pixel convolution was proposed as the final layer of deep SR network \cite{dong2016accelerating}. However, using only low-frequency features to train the SR model may result in unstable training \cite{ahn2018image}. In terms of this issue, we set the sub-pixel into the reconstruction block, RB as the middle part of the LESRCNN to covert low-frequency features into high-frequency features, then, the output of the RB acts the IRB, where the IRB can further learn more robust high-frequency features. Although enhancement operation in the IEEB can make the information of shallow layers transmit the deep layers, up-sampling operation may loss some information of original LR image. To deal with this problem, we propose a two-step mechanism in the RB. The first step uses the sub-pixel convolution technique to upsample the outputs of the IEEB and the first layer of the IEEB as the local and global features, respectively. The second step utilizes the RL method to gather local and global features. The two-step mechanism above can further improve the expressive ability for a SISR model. Also, that is complementary to the IEEB in addressing the long-term dependency problem. The effectiveness of the two-step RB is tested by the Table 1, where `IEEB+RB' obtains better results of PSNR and SSIM than that of the `IEEB' on the Set5.

(3) Information refinement block: According to Section 2.1, the combination of using low- and high-frequency features can achieve notable improvement over single technique. However, the IEEB emphasizes the effects of low-frequency features and the RB has power ability of converting low-frequency features into coarse high-frequency features, which ignores the influences of high-frequency features. Motivated by that, an information refinement block, IRB is developed. The IRB with four Conv+ReLU can learn more accurate high-frequency features via coarse high-frequency features from the RB, which can reduce the difference between the predicted SR image and the target HR image. Additionally, it can reconstruct a SR image by a single convolution. The IRB has great improvement on performance for the LESRCNN as shown in Table 1, where the `LESRCNN' outperforms `IEEB+RB' in both of PSNR and SSIM.
\subsection{Comparisons with state-of-the-arts}
To better evaluate the results of the LESRCNN in SISR, both of quantitative and qualitative analysis are chosen. The quantitative analysis depends on both of PSNR and SSIM, running time of a LR image and complexities of the popular methods, i.e., Bicubic, A+ \cite{timofte2014a+}, jointly optimized regressors (JOR) \cite{dai2015jointly}, RFL \cite{schulter2015fast}, self-exemplars super-resolution (SelfEx) \cite{huang2015single}, CSCN \cite{wang2015deep}, RED \cite{mao2016image}, a denoising convolutional neural network (DnCNN) \cite{zhang2017beyond}, trainable nonlinear reaction diffusion (TNRD) \cite{chen2016trainable}, fast dilated residual SR convolutional network (FDSR) \cite{lu2018fast}, SRCNN \cite{dong2015image}, FSRCNN \cite{dong2016accelerating}, residue context sub-network (RCN) \cite{shi2017structure}, VDSR \cite{kim2016accurate}, deeply-recursive convolutional network (DRCN) \cite{kim2016deeply}, context-wise network fusion (CNF) \cite{ren2017image}, Laplacian SR network (LapSRN) \cite{lai2017deep}, MemNet \cite{tai2017memnet}, CARN-M \cite{ahn2018fast}, wavelet domain residual network (WaveResNet) \cite{bae2017beyond}, convolutional principal component analysis (CPCA) \cite{xu2018self}, new architecture of deep recursive convolution networks for SR (NDRCN) \cite{cao2019new}, LESRCNN and LESRCNN-S on four benchmark datasets (e.g. Set5, Set14, B100 and U100) for SISR. The qualitative analysis is explained by visual figures. Further, we introduce the quantitative and qualitative evaluation as follows.

Both of PSNR and SSIM are expressed through Tables 4-7. From the Table 4, we can see that the proposed LESRCNN and LESRCNN-S can obtain superior performance against state-of-the-art SR methods with scale factors of $\times 3$ and $\times 4$ on Set5 for SISR, respectively, where the LESRCNN denotes a SR model for a certain scale and the LESRCNN-S is a SR model for three scales (i.e., $\times 2$, $\times 3$ and $\times 4$). The LESRCNN-S is very suitable to real applications. Also, the LESRCNN achieves the similar result to the CNF in PSRN for $\times 2$  on the Set5. The LESRCNN-S is 0.19dB higher than the WaveResNet for $\times 3$ in Table 4. The LESRCNN obtains the best performance with three different scale factors, such as $\times 2$, $\times 3$ and $\times 4$ on the Set14 in SISR as illustrated in Table 5. For example, the LESRCNN obtains the improvements in PSNR of 0.04 dB and SSIM of 0.0006 than that of the popular methods, i.e., MemNet for scale factor of $\times 2$ on the Set14.

The LESRCNN is suitable to large-scale datasets, such as B100 and U100. According to the Tables 6 and 7, we can see that the proposed LESRCNN has obvious superiority than that of other popular methods, such as the CARN-M. For example, the LESRCNN has obtained notable gain both of PSRN of 0.22dB and SSIM of 0.0013 in contrast to the CARN-M for scale factor of $\times 2$ on the U100 as shown in Table 7.  Additionally, the LESRCNN-S is also very competitive to the popular SR methods on B100 and U100. For instance, the LESRCNN-S can achieve the improvements in PSNR of 0.15dB and SSIM of 0.0045 than that of the CARN-M for $\times 4$ on U100. According to the results, we find that the LESRCNN and LESRCNN-S perform well for SR task.
\begin{table}[t!]
\caption{PSNR and SSIM of different techniques with scale factors of $\times 2$, $\times 3$ and $\times 4$ on Set5.}
\label{tab:1}
\centering
\scalebox{1}[1]{
\begin{tabular}{|c|c|c|c|c|}
\hline
\multirow{2}{*}{Dataset} &
\multirow{2}{*}{Model} &
$\times 2$ & $\times 3$ & $\times 4$\\
\cline{3-5} & &PSNR/SSIM &PSNR/SSIM &PSNR/SSIM\\
\hline
\multirow{18}{*}{Set5} &
Bicubic	&33.66/0.9299	&30.39/0.8682	&28.42/0.8104\\
\cline{2-5} &
A+ \cite{timofte2014a+}	&36.54/0.9544	&32.58/0.9088	&30.28/0.8603\\
\cline{2-5} &
JOR \cite{dai2015jointly}	&36.58/0.9543	&32.55/0.9067	&30.19/0.8563\\
\cline{2-5} &
RFL \cite{schulter2015fast}	&36.54/0.9537	&32.43/0.9057	&30.14/0.8548\\
\cline{2-5} &
SelfEx \cite{huang2015single}	&36.49/0.9537	&32.58/0.9093	&30.31/0.8619\\
\cline{2-5} &
CSCN \cite{wang2015deep}	&36.93/0.9552	&33.10/0.9144	&30.86/0.8732\\
\cline{2-5} &
RED \cite{mao2016image}	&37.56/0.9595	&33.70/0.9222	&31.33/0.8847\\
\cline{2-5} &
DnCNN \cite{zhang2017beyond}	&37.58/0.9590	&33.75/0.9222	&31.40/0.8845\\
\cline{2-5} &
TNRD \cite{chen2016trainable}	&36.86/0.9556	&33.18/0.9152	&30.85/0.8732\\
\cline{2-5} &
FDSR \cite{lu2018fast}	&37.40/0.9513	&33.68/0.9096	&31.28/0.8658\\
\cline{2-5} &
SRCNN \cite{dong2015image}	&36.66/0.9542	&32.75/0.9090	&30.48/0.8628\\
\cline{2-5} &
FSRCNN \cite{dong2016accelerating}	&37.00/0.9558	&33.16/0.9140	&30.71/0.8657\\
\cline{2-5} &
RCN \cite{shi2017structure}	&37.17/0.9583	&33.45/0.9175	&31.11/0.8736\\
\cline{2-5} &
VDSR \cite{kim2016accurate}	&37.53/0.9587	&33.66/0.9213	&31.35/0.8838\\
\cline{2-5} &
DRCN \cite{kim2016deeply}	&37.63/0.9588	&33.82/0.9226	&31.53/0.8854\\
\cline{2-5} &
CNF \cite{ren2017image}	&37.66/0.9590	&33.74/0.9226	&31.55/0.8856\\
\cline{2-5} &
LapSRN \cite{lai2017deep}	&37.52/0.9590	&-	&31.54/0.8850 \\
\cline{2-5} &
MemNet \cite{tai2017memnet} &37.78/0.9597 &34.09/0.9248 &31.74/0.8893\\
\cline{2-5} &
CARN-M \cite{ahn2018fast} &37.53/0.9583 &33.99/0.9236 &31.92/0.8903\\
\cline{2-5}&
WaveResNet \cite{bae2017beyond}	&37.57/0.9586	&33.86/0.9228	&31.52/0.8864\\
\cline{2-5} &
CPCA \cite{xu2018self}	&34.99/0.9469	&31.09/0.8975	&28.67/0.8434\\
\cline{2-5} &
NDRCN \cite{cao2019new} &37.73/0.9596 &33.90/0.9235 &31.50/0.8859\\
\cline{2-5} &
LESRCNN (Ours)	&37.65/0.9586	&33.93/0.9231	&31.88/0.8903\\
\cline{2-5} &
LESRCNN-S (Ours) &37.57/0.9582 &34.05/0.9238 &31.88/0.8907\\
\hline
\end{tabular}}
\label{tab:booktabs}
\end{table}
\begin{table}[t!]
\caption{PSNR and SSIM of different techniques with scale factors of $\times 2$, $\times 3$ and $\times 4$ on Set14.}
\label{tab:1}
\centering
\scalebox{1}[1]{
\begin{tabular}{|c|c|c|c|c|}
\hline
\multirow{2}{*}{Dataset} &
\multirow{2}{*}{Model} &
$\times 2$ & $\times 3$ & $\times 4$\\
\cline{3-5} & &PSNR/SSIM &PSNR/SSIM &PSNR/SSIM\\
\hline
\multirow{20}{*}{Set14} &
Bicubic	&30.24/0.8688	&27.55/0.7742	&26.00/0.7027\\
\cline{2-5} &
A+ \cite{timofte2014a+}	&32.28/0.9056	&29.13/0.8188	&27.32/0.7491\\
\cline{2-5} &
JOR \cite{dai2015jointly}	&32.38/0.9063	&29.19/0.8204	&27.27/0.7479\\
\cline{2-5} &
RFL \cite{schulter2015fast}	&32.26/0.9040	&29.05/0.8164	&27.24/0.7451\\
\cline{2-5} &
SelfEx \cite{huang2015single}	&32.22/0.9034 &29.16/0.8196	&27.40/0.7518\\
\cline{2-5} &
CSCN \cite{wang2015deep}	&32.56/0.9074	&29.41/0.8238 &27.64/0.7578\\
\cline{2-5} &
RED \cite{mao2016image}	&32.81/0.9135	&29.50/0.8334	&27.72/0.7698\\
\cline{2-5} &
DnCNN \cite{zhang2017beyond}	&33.03/0.9128	&29.81/0.8321	&28.04/0.7672\\
\cline{2-5} &
TNRD \cite{chen2016trainable}	&32.51/0.9069	&29.43/0.8232	&27.66/0.7563\\
\cline{2-5} &
FDSR \cite{lu2018fast}	&33.00/0.9042	&29.61/0.8179	&27.86/0.7500\\
\cline{2-5} &
SRCNN \cite{dong2015image}	&32.42/0.9063	&29.28/0.8209	&27.49/0.7503\\
\cline{2-5} &
FSRCNN \cite{dong2016accelerating}	&32.63/0.9088	&29.43/0.8242	&27.59/0.7535\\
\cline{2-5} &
RCN \cite{shi2017structure}	&32.77/0.9109	 &29.63/0.8269	&27.79/0.7594\\
\cline{2-5} &
VDSR \cite{kim2016accurate}	&33.03/0.9124	&29.77/0.8314 &28.01/0.7674\\
\cline{2-5} &
DRCN \cite{kim2016deeply}	&33.04/0.9118	&29.76/0.8311	&28.02/0.7670\\
\cline{2-5} &
CNF \cite{ren2017image}	&33.38/0.9136	&29.90/0.8322	&28.15/0.7680\\
\cline{2-5} &
LapSRN \cite{lai2017deep}	&33.08/0.9130	&29.63/0.8269	&28.19/0.7720\\
\cline{2-5} &
MemNet \cite{tai2017memnet}	&33.28/0.9142	&30.00/0.8350	&28.26/0.7723\\
\cline{2-5} &
CARN-M \cite{ahn2018fast}	&33.26/0.9141	&30.08/0.8367	&28.42/0.7762\\
\cline{2-5} &
WaveResNet \cite{bae2017beyond}	&33.09/0.9129	 &29.88/0.8331	&28.11/0.7699\\
\cline{2-5} &
CPCA \cite{xu2018self}	&31.04/0.8951	&27.89/0.8038	&26.10/0.7296\\
\cline{2-5} &
NDRCN \cite{cao2019new} &33.20/0.9141 &29.88/0.8333 &28.10/0.7697\\
\cline{2-5} &
LESRCNN (Ours)	&33.32/0.9148	&30.12/0.8380	&28.44/0.7772\\
\cline{2-5} &
LESRCNN-S (Ours) &33.30/0.9145 &30.16/0.8384 &28.43/0.7776\\
\hline
\end{tabular}}
\label{tab:booktabs}
\end{table}
\begin{table}[t!]
\caption{PSNR and SSIM of different techniques with scale factors of $\times 2$, $\times 3$ and $\times 4$ on B100.}
\label{tab:1}
\centering
\scalebox{1}[1]{
\begin{tabular}{|c|c|c|c|c|}
\hline
\multirow{2}{*}{Dataset} &
\multirow{2}{*}{Model} &
$\times 2$ & $\times 3$ & $\times 4$\\
\cline{3-5} & &PSNR/SSIM &PSNR/SSIM &PSNR/SSIM\\
\hline
\multirow{17}{*}{B100} &
Bicubic	&29.56/0.8431	&27.21/0.7385	&25.96/0.6675\\
\cline{2-5} &
A+ \cite{timofte2014a+}	&31.21/0.8863	&28.29/0.7835	&26.82/0.7087\\
\cline{2-5} &
JOR \cite{dai2015jointly}	&31.22/0.8867	&28.27/0.7837	&26.79/0.7083\\
\cline{2-5} &
RFL \cite{schulter2015fast}	&31.16/0.8840	&28.22/0.7806	&26.75/0.7054\\
\cline{2-5} &
SelfEx \cite{huang2015single}	&31.18/0.8855	&28.29/0.7840	&26.84/0.7106\\
\cline{2-5} &
CSCN \cite{wang2015deep}	&31.40/0.8884	&28.50/0.7885	&27.03/0.7161\\
\cline{2-5} &
RED \cite{mao2016image}	&31.96/0.8972	&28.88/0.7993	&27.35/0.7276\\
\cline{2-5} &
DnCNN \cite{zhang2017beyond}	&31.90/0.8961	&28.85/0.7981	&27.29/0.7253\\
\cline{2-5} &
TNRD \cite{chen2016trainable}	&31.40/0.8878	&28.50/0.7881	&27.00/0.7140\\
\cline{2-5} &
FDSR \cite{lu2018fast}	&31.87/0.8847	&28.82/0.7797	&27.31/0.7031\\
\cline{2-5} &
SRCNN \cite{dong2015image}	&31.36/0.8879	&28.41/0.7863	&26.90/0.7101\\
\cline{2-5} &
FSRCNN \cite{dong2016accelerating}	&31.53/0.8920	&28.53/0.7910	&26.98/0.7150\\
\cline{2-5} &
VDSR \cite{kim2016accurate}	&31.90/0.8960	&28.82/0.7976	&27.29/0.7251\\
\cline{2-5} &
DRCN \cite{kim2016deeply}	&31.85/0.8942	&28.80/0.7963	&27.23/0.7233\\
\cline{2-5} &
CNF \cite{ren2017image}	&31.91/0.8962	&28.82/0.7980	&27.32/0.7253\\
\cline{2-5} &
LapSRN \cite{lai2017deep}	&31.80/0.8950	&-	&27.32/0.7280\\
\cline{2-5} &
MemNet \cite{tai2017memnet}	&32.08/0.8978 &28.96/0.8001 &27.40/0.7281\\
\cline{2-5}&
CARN-M \cite{ahn2018fast}	&31.92/0.8960	&28.91/0.8000	&27.44/0.7304\\
\cline{2-5} &
WaveResNet \cite{bae2017beyond} &32.15/0.8995 &28.86/0.7987 &27.32/0.7266\\
\cline{2-5} &
NDRCN \cite{cao2019new} &32.00/0.8975 &28.86/0.7991 &27.30/0.7263\\
\cline{2-5} &
LESRCNN (Ours)	&31.95/0.8964	&28.91/0.8005	&27.45/0.7313\\
\cline{2-5} &
LESRCNN-S (Ours) &31.95/0.8965 &28.94/0.8012 &27.47/0.7321\\
\hline
\end{tabular}}
\label{tab:booktabs}
\end{table}
\begin{table}[t!]
\caption{PSNR and SSIM of different techniques with scale factors of $\times 2$, $\times 3$ and $\times 4$ on U100.}
\label{tab:1}
\centering
\scalebox{1}[1]{
\begin{tabular}{|c|c|c|c|c|}
\hline
\multirow{2}{*}{Dataset} &
\multirow{2}{*}{Model} &
$\times 2$ & $\times 3$ & $\times 4$\\
\cline{3-5} & &PSNR/SSIM &PSNR/SSIM &PSNR/SSIM\\
\hline
\multirow{18}{*}{U100} &
Bicubic	&26.88/0.8403	&24.46/0.7349	&23.14/0.6577\\
\cline{2-5} &
A+ \cite{timofte2014a+}	&29.20/0.8938	&26.03/0.7973	&24.32/0.7183\\
\cline{2-5} &
JOR \cite{dai2015jointly}	&29.25/0.8951	&25.97/0.7972	&24.29/0.7181\\
\cline{2-5} &
RFL \cite{schulter2015fast}	&29.11/0.8904	&25.86/0.7900	&24.19/0.7096\\
\cline{2-5} &
SelfEx \cite{huang2015single}	&29.54/0.8967	&26.44/0.8088 &24.79/0.7374\\
\cline{2-5} &
DnCNN \cite{zhang2017beyond}	&30.74/0.9139	&27.15/0.8276	&25.20/0.7521\\
\cline{2-5} &
TNRD \cite{chen2016trainable} &29.70/0.8994	&26.42/0.8076	&24.61/0.7291\\
\cline{2-5} &
FDSR \cite{lu2018fast} &30.91/0.9088	&27.23/0.8190	&25.27/0.7417\\
\cline{2-5} &
SRCNN \cite{dong2015image} &29.50/0.8946	&26.24/0.7989	&24.52/0.7221\\
\cline{2-5} &
FSRCNN \cite{dong2016accelerating} &29.88/0.9020	&26.43/0.8080	&24.62/0.7280\\
\cline{2-5} &
VDSR \cite{kim2016accurate} &30.76/0.9140	&27.14/0.8279	&25.18/0.7524\\
\cline{2-5} &
DRCN \cite{kim2016deeply} &30.75/0.9133	&27.15/0.8276	&25.14/0.7510\\
\cline{2-5} &
LapSRN \cite{lai2017deep} &30.41/0.9100	&-	&25.21/0.7560\\
\cline{2-5} &
MemNet \cite{tai2017memnet}	&31.31/0.9195	&27.56/0.8376	&25.50/0.7630\\
\cline{2-5} &
CARN-M \cite{ahn2018fast}	&31.23/0.9193	&27.55/0.8385	&25.62/0.7694\\
\cline{2-5} &
WaveResNet \cite{bae2017beyond}	&30.96/0.9169	 &27.28/0.8334	 &25.36/0.7614\\
\cline{2-5} &
CPCA \cite{xu2018self}	&28.17/0.8990	&25.61/0.8123	&23.62/0.7257\\
\cline{2-5} &
NDRCN \cite{cao2019new} &31.06/0.9175 &27.23/0.8312 &25.16/0.7546\\
\cline{2-5} &
LESRCNN (Ours)	&31.45/0.9206	&27.70/0.8415	&25.77/0.7732\\
\cline{2-5} &
LESRCNN-S (Ours) &31.45/0.9207 &27.76/0.8424 &25.78/0.7739\\
\hline
\end{tabular}}
\label{tab:booktabs}
\end{table}

For running time, we choose four methods to test the running time of an image with different sizes (i.e., $256 \times 256$, $512 \times 512$ and $1024 \times 1024$). As explained in Table 8, it is known that the LESRCNN has faster execution to deal with the images of three different sizes than that of VDSR, MemNet and CARN-M.
\begin{table}[t!]
\caption{Running time of four networks for recovering the images of sizes $256\times256$, $512\times512$ and $1024\times1024$.}
\label{tab:1}
\centering
\scalebox{1}[1]{
\begin{tabular}{|c|c|c|c|c|}
\hline
\multicolumn{5}{|c|}{Single Image Super-Resolution} \\
\hline
Size &VDSR \cite{kim2016accurate}  &MemNet \cite{tai2017memnet} &CARN-M \cite{ahn2018fast}  &LESRCNN (Ours)\\
\hline
$256 \times 256$ &0.0172	&0.8774 &0.0159 &0.0102\\
\hline
$512 \times 512$ &0.0575	&3.605	&0.0199	&0.0129\\
\hline
$1024 \times 1024$ &0.2126	&14.69 &0.0320	&0.0222\\
\hline
\end{tabular}}
\label{tab:booktabs}
\end{table}
\begin{table}[t!]
\caption{Complexity of five networks for SISR.}
\label{tab:1}
\centering
\scalebox{1}[1]{
\begin{tabular}{|c|c|c|}
\hline
Methods	&Parameters	&Flops\\
\hline
VDSR \cite{kim2016accurate}	&665K	&10.90G\\
\hline
DnCNN \cite{zhang2017beyond}	&556K	&9.18G\\
\hline
DRCN \cite{kim2016deeply}	&1774K	&29.07G\\
\hline
MemNet \cite{tai2017memnet}	&677K	&11.09G\\
\hline
LESRCNN	(Ours) &516K	&3.08G\\
\hline
\end{tabular}}
\label{tab:booktabs}
\end{table}
\begin{figure}[!htb]
\centering
\subfloat{\includegraphics[width=6.5in]{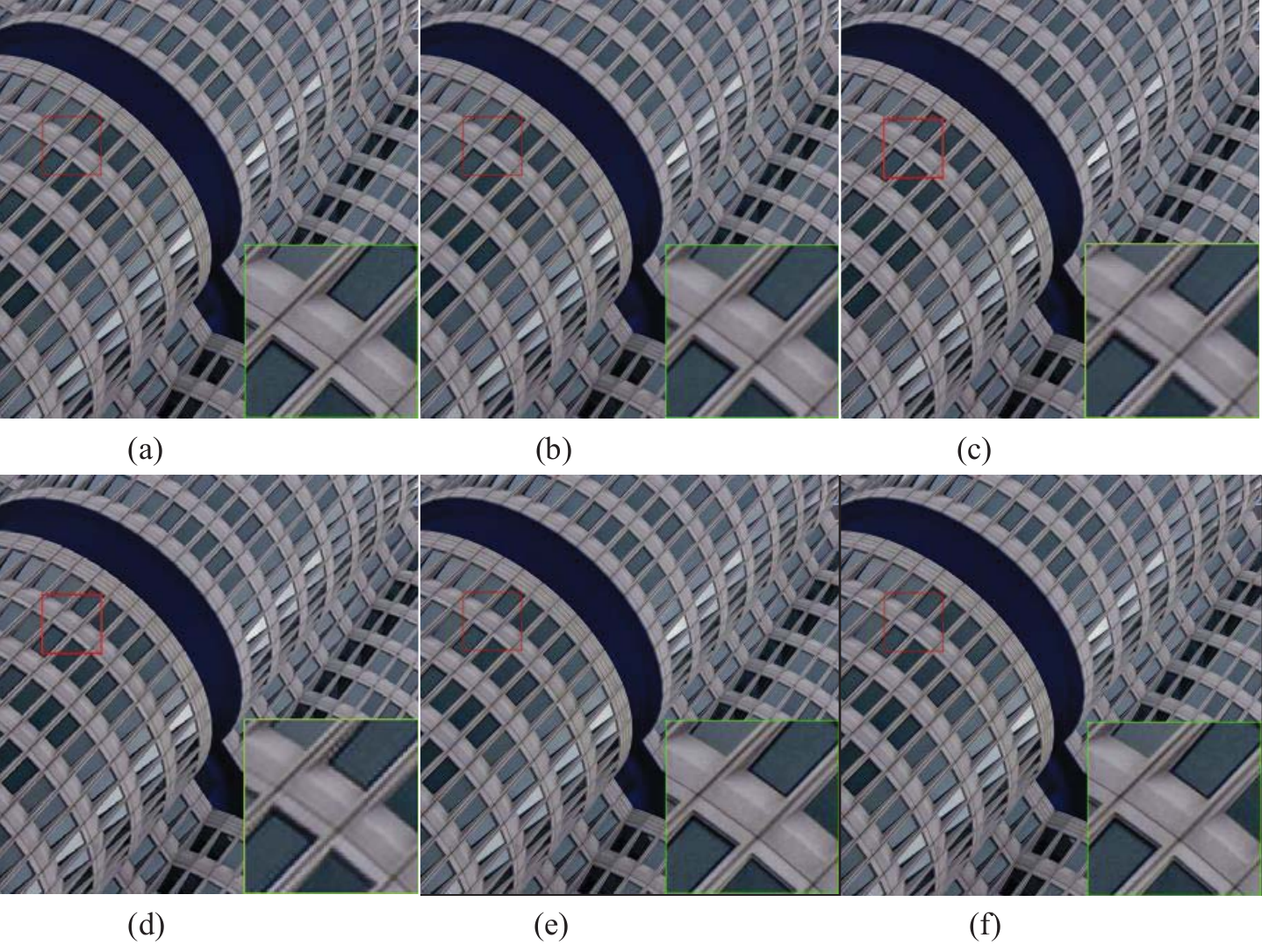}
\label{fig_second_case}}
\caption{Visual comparisons of different methods on an image from the U100 for $\times 2$ scale: (a) A HR image (PSNR/SSIM), (b)Bicubic (31.88/0.9187), (c) SelfEx (32.92/0.9413), (d) SRCNN (31.88/0.9259), (e) CARN-M (37.59/0.9738) and (f) LESRCNN (37.95/0.9746).}
\label{fig:5}
\end{figure}

In terms of the complexity, five methods are utilized to test the number of parameters and flops. From the Table 9, we can see that the LESRCNN uses less parameters and flops than that of the state-of-the-art SR techniques, such as MemNet, which indicates the LESRCNN has lower computational cost and less memory consumption for training phase. In a summary, the proposed LESRCNN is superior to other state-of-the-art SR methods, such as MemNet and CARN-M in quantitative analysis.
\begin{figure}[!htb]
\centering
\subfloat{\includegraphics[width=6.5in]{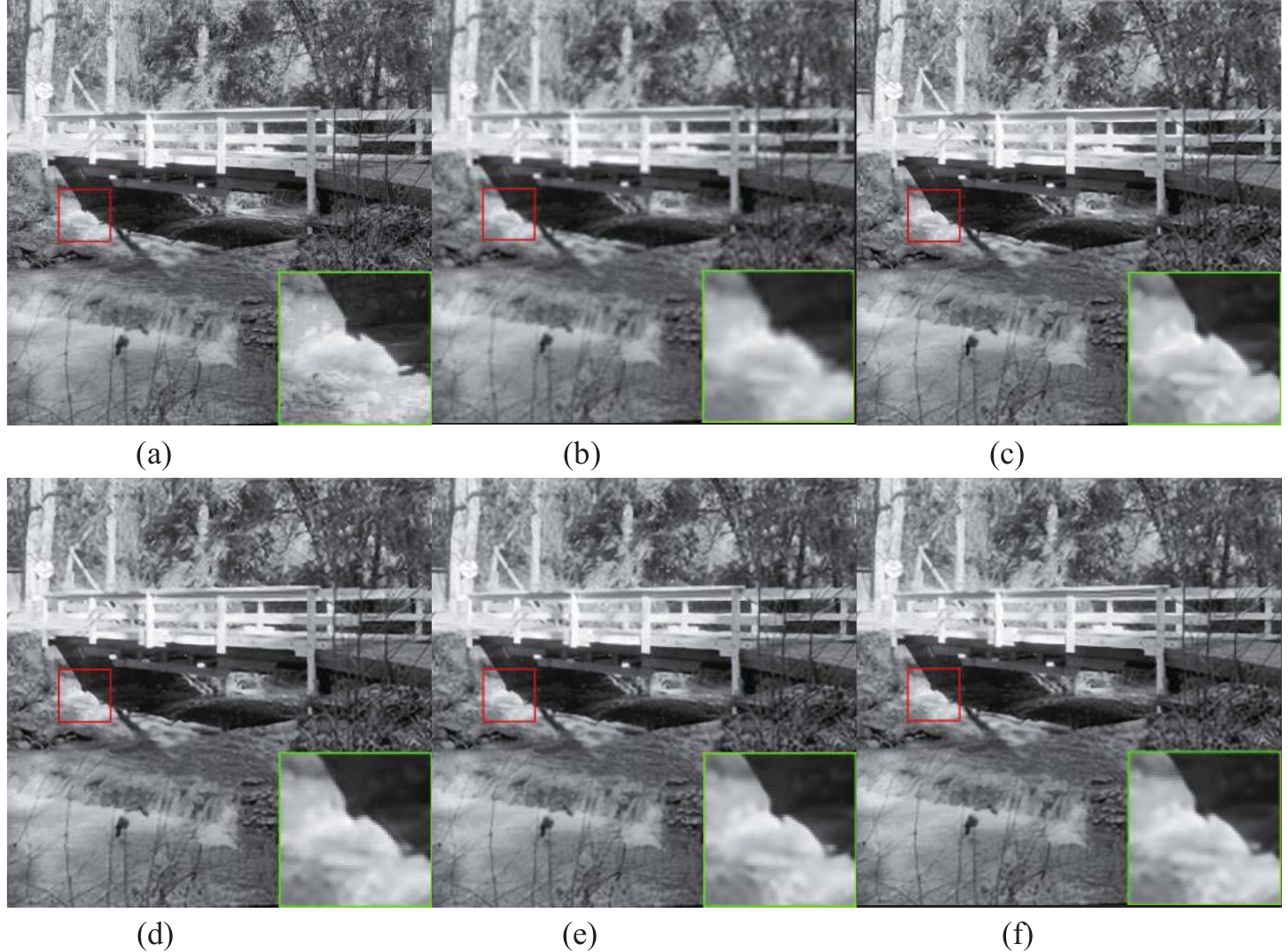}
\label{fig_second_case}}
\caption{Visual results of different methods on an image from the Set14 for $\times 3$ scale: (a) A HR image (PSNR/SSIM), (b)Bicubic (25.65/0.6658), (c) SelfEx (26.03/0.7086), (d) SRCNN (25.92/0.7033), (e) CARN-M (26.34/0.7856) and (f) LESRCNN (26.82/0.7948).}
\label{fig:5}
\end{figure}

For qualitative analysis, we use six visual figures from the given HR image, Bicubic, SelfEx, SRCNN, CARN-M and LESRCNN to test the effects of the predicted SR image for $\times 2$, $\times 3$ and $\times 4$, respectively. As shown in Figs. 3-5, we can see that the magnified area of the predicted SR image from the LESRCNN is clearer than other methods, such as CARN-M for three different scales. That shows that our proposed LESRCNN is competitive in qualitative evaluation. According to the quantitative and qualitative analysis above, the LESRCNN is more effective for SISR.
\begin{figure}[!htb]
\centering
\subfloat{\includegraphics[width=6.5in]{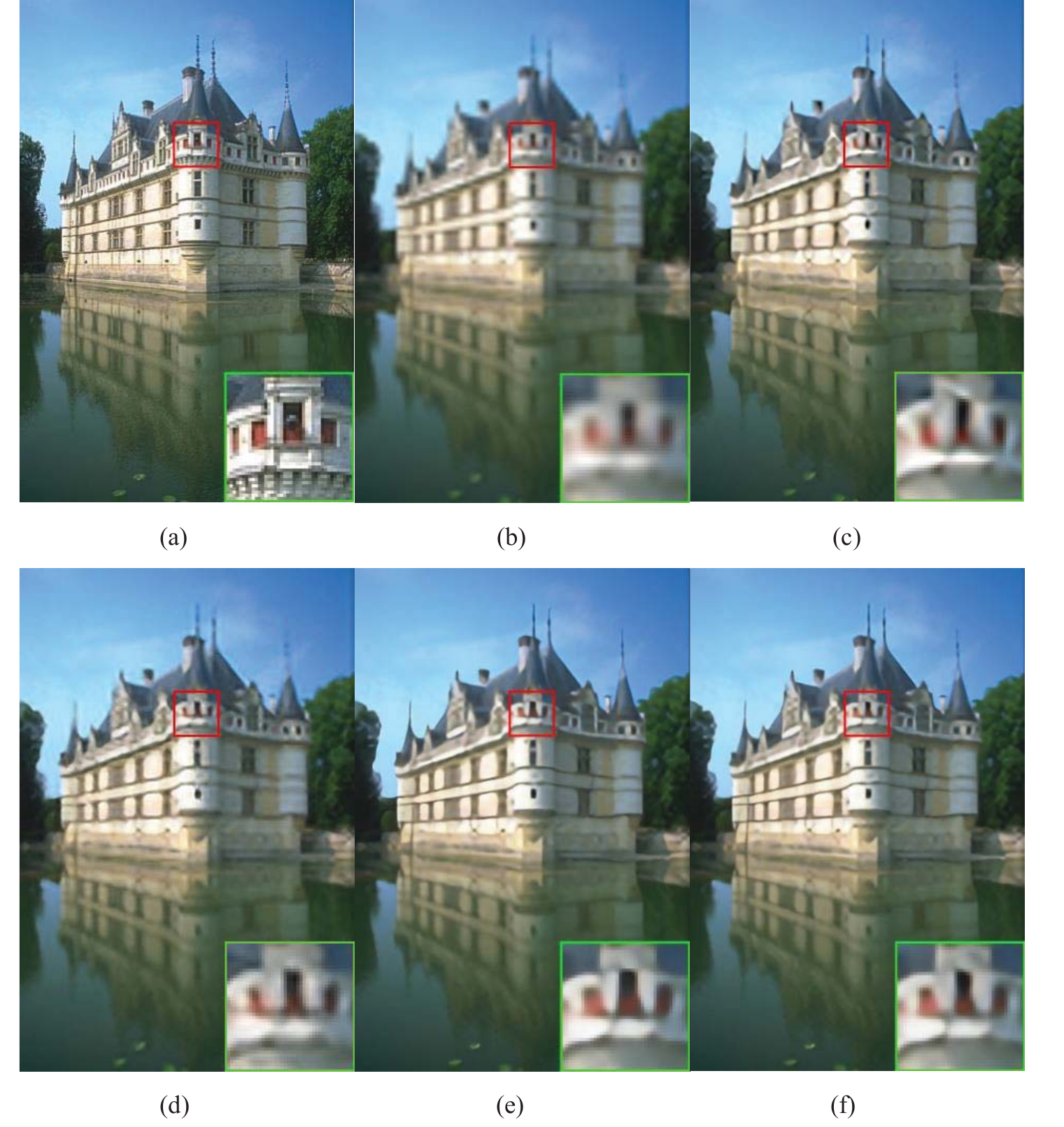}
\label{fig_second_case}}
\caption{Visual effects of different methods on an image from the B100 for $\times 4$ scale: (a) A HR image (PSNR/SSIM), (b) Bicubic (25.26/0.7539), (c) SelfEx (25.83/0.7852), (d) SRCNN (25.78/0.7767), (e) CARN-M (26.39/0.8046) and (f) LESRCNN (26.46/0.8061).}
\label{fig:5}
\end{figure}
\section{Conclusion}
In this paper, we propose a lightweight enhanced super-resolution CNN (LESRCNN) by cascading an IEEB, a RB and an IRB. The IEEB can extract and aggregate hierarchical low-frequency features, addressing the long-term dependency problem. Also, a heterogeneous architecture is fused into the IEEB to reduce the number of parameters and complexity for training a SR model. The RB can convert low-frequency features into high-frequency features by fusing global and local features, which is complementary with the IEEB in enhancing the memory ability of shallow layers on deep layers in SISR. The IRB uses coarse high-frequency features from the RB to learn more accurate SR features and construct a SR image. The LESRCNN obtains a high-resolution image via a model and multiple models for different scales. Extensive experiments illustrate that the proposed LESRCNN outperforms state-of-the-arts on SISR in terms of qualitative and quantitative evaluation.
\\
\\
\\
\section*{Acknowledgments}
C. Tian, R. Zhu, Z. Wu and Y. Xu are supported in part by the National Nature Science Foundation of China Gant No. 61876051, and in part by the the  Shenzhen Key  Laboratory  of  Visual  Object  Detection  and  Recognition  under  Grant No. ZDSYS20190902093015527 in this work.
%\section*{References}
%% References


% \bibliographystyle{elsarticle-num}
\bibliographystyle{elsarticle-harv}
\bibliography{references}
%\bibliography{sample}

%-----------------------------------------------------------------------------------------------------
\end{spacing}
\end{document}